\makeatletter
\documentclass[twocolumn]{aastex631}
\usepackage{listings}
\usepackage{subfigure}
\usepackage{xcolor}
\usepackage{color,soul}
\usepackage{enumitem}
\hypersetup{backref=true, pagebackref=true, hyperindex=true, colorlinks=true,                
	breaklinks=true, urlcolor= magenta, linkcolor= black, bookmarks=true,                 
	bookmarksopen=false, filecolor=cyan, citecolor=blue, linkbordercolor=blue}
\usepackage{hyperref}

\newcommand{\gaia}{\textit{Gaia}}

\shorttitle{Mass-Age-Velocity Dispersion Relation}
\shortauthors{Fatheddin and Sajadian}

\usepackage{amsmath}
\begin{document}
\title{A Statistical Relation between Mass, Age and Velocity Dispersion in the Solar Neighborhood \footnote{Based on the third data release of the \textit{Gaia} Space Telescope}}

\author[0000-0002-7611-9249]{Hossein Fatheddin}
\affiliation{Department of Physics, Isfahan University of Technology, Isfahan 84156-83111, Iran}
 
\author[0000-0002-0167-3595]{Sedighe Sajadian}
\affiliation{Department of Physics, Isfahan University of Technology, Isfahan 84156-83111, Iran}

\begin{abstract}
The stellar kinematics of the Galactic disk are main factors for constraining disk formation and evolution processes in the Milky Way (MW) Galaxy. In this paper we investigate a statistical relation between stellar Mass, Age and Velocity Dispersion for stars in the Solar neighborhood. Age-Velocity dispersion relations (AVR), with their applications, have been studied in details before. But their correlation with mass was mostly neglected. To investigate this relation, we use the proper motion data of more than 113035 stars in the Galactic disk (with Solar distances less than 150 parsecs) provided by the third data release of the \textit{Gaia} mission and for stellar Mass and Age, \textit{Gaia}'s Final Luminosity Age Mass Estimator (\textit{FLAME}) is implemented. We analyze this data and the correlations between the parameters with Random Forrest (RF) Regression, which is an Ensemble statistical learning technique. Finally, we show that by considering the stellar mass alongside age, we can determine Velocity Dispersions with the average relative error, and mean absolute error of about $9\%$, and $2.68~\rm{km/s}$, respectively. We also find that the correlation of stellar age with Velocity Dispersion is $3$ to $8$ times more than mass, which varies due to the different stellar types and masses.
\end{abstract}
\keywords{Astrostatistics techniques(1886) --- Stellar dynamics(1596) --- Solar neighborhood(1509)}

\section{Introduction} \label{sec:intro}
In our Galaxy, stars rotate around the Galactic center due to the global gravitational potential of the Galaxy. The global rotational velocity is a function of the distance from the Galactic center \citep[see, e.g., ][]{Brand1993}. In addition to this global velocity, there is another velocity component for stars in the Galaxy which is random (or peculiar) velocities. This velocity for each given star can be in any direction and is generally defined in three components, i.e., $v_{\rm U},~v_{\rm V},~v_{\rm W}$, where U is toward the Galactic center, V is in the direction of Galactic rotation and W is normal to the Galactic plane and toward the north Galactic pole \citep[see, e.g.,][]{2010Schonrich}.

\noindent The random velocities are mostly caused by the gravitational interaction of stars with the Galactic giant structures and are not an intrinsic property of the star. Since the dynamical friction in the gas that originated the star, tends to circularize the star's orbit \citep{Bonetti2020}. For instance, the Galactic spiral arms, Galactic bar, giant gas or molecular clouds, globular clusters, orbiting satellite galaxies and galaxy mergers can encounter stars and generally cause a disk heating or cooling \citep[see, e.g., ][]{1967Barbanis,1968Woolley,2008BinnyTBook,2009Quillen}. These interactions cause dependence of random velocities with either stellar locations in the Galactic disk, or physical parameters of stars \citep[see, e.g., ][]{2016ApJSumi,sajadian2019v}. They are briefly explained in the following items. 

\setlist[itemize]{nosep,topsep={-5pt},partopsep={-5pt}}
\begin{itemize}[leftmargin=2.0mm]
\item  The Galactic transient spiral arms have asymmetric gravitational potentials which change with time. The transient giant patterns cause disk heating, and as a result, oscillations mostly in the radial and rotational directions (inside the Galactic disk) \citep{1967Barbanis}. These structures can scatter stars and gas materials even around the Lindblad resonance radii \citep{1980Goldreich}, where they potentially can emigrate from their initial birth orbits \citep{SellwoodBenney2002}. Because of such encounters, the gas clouds inside the Galactic disk also have non-circular orbits around the Galactic center \citep{2003Bissantz}. Based on numerical integrations, the encounter with the transient spiral patterns causes the change in radial random velocity of stars with time as $v_{\rm U}\propto  t^{0.3}$, which was well supported by the observations \citep{2004DeSimone}.
\item  Massive gas and molecular clouds might be accountable for stellar scattering in the Galactic disk \citep{1953Spitzer}. A gas cloud in the Galactic disk will make epicycle motions for the stellar radial and rotational velocities, and redistribute these components even to the vertical one \citep{2008BinnyTBook}.
\item Additionally, The Galactic bar has considerable effects on the disk stars' velocity, specially in the vertical direction \citep{2000AJDehnen,2010Minchev}. 
\item Galactic mergers can also cause extensive disk heating and make significant perturbations to the kinematic structure of the disk. \citep{Quinn1993}
\item MAssive Compact Halo Objects (MACHOs) in the Galactic halo can encounter with the disk stars which causes both vertical and radial random velocities increase with time as $t^{0.5}$, with their ratio as $0.53$ \citep{1985Lacey,2008BinnyTBook}.\\
\end{itemize}

Because of these mentioned interactions, the stellar random velocities evolve and increase with time, and older stars are further affected by such heatings. As a result, we expect that older stars have higher random velocities, i.e., the known Age-Velocity Dispersion Relation, which was confirmed through different survey observations as well \citep{Gomez1997,Nordstrom2004,Jincheng2018,Mackereth2019}.  

The main heating mechanisms in the Galactic disk (i.e., Spiral arms and gas clouds) offer similar heating process for all stars with different masses.  For that reason, we do not expect any correlation between stellar mass and AVR at first sight. However, some second-order interactions (i.e., stellar encounter with each others or MACHOs) cause mass-dependent evolutions in random velocities. By considering their time scales which should be larger than the age of our Galaxy, we expect a weak mass-dependent relation. Hence, a large set of observing data from stellar kinematic measurements and astrophysical parameters might resolve such weak and second-order correlation between mass and AVR for stars in the Galactic disk.  

\textit{Gaia Space Observatory} has provided (and provides) such accurate data sets which enable us to study this correlation. In \citet{Sajadian2022}, we have used the \textit{Gaia}~data archive and shown that there is a correlation between the stellar random velocity and mass at $2$-$3$ sigma level, however by averaging over the stellar age. In that paper, we made 10 subsamples of \textit{Gaia}~data versus mass and for each subsample estimated the scale parameters, $a$, of the best-fitted Maxwell-Boltzmann distributions to random velocities. In fact there was a linear relation between $a$ and the inverse of square root of average mass.  

In this work, we will extend the previous work and try to find a three-fold correlation between mass, age, and velocity dispersion of stars in the \textit{Gaia}~data. For this, We implement a powerful Statistical technique called Random Forrest which belongs to a class of classical supervised learning methods. 

The outline of the paper is as follows. In Section \ref{sec:Data}, we will describe the \gaia~data used in this paper, and all the filters that are applied to the data. Our methods for preparing and engineering the data before modeling and analysis, which are done in Section \ref{sec:Method}, are also explained in \ref{sec:Data}. we explain the results of our analysis and the three-fold correlation in Section \ref{sec:Results}. In the last section, \ref{Conclusions}, we summarize the results.  

\section{Data} \label{sec:Data}
The third data release of the \textit{Gaia Space Observatory}, which is available through the \textit{Gaia} Archive \footnote{\url{https://gea.esac.esa.int/archive/}}, contains the full astrometric solutions for around 1.46 billion sources, with a limiting magnitude of about $G \approx 21$ and a bright limit of about $G \approx 3$ \citep{GaiaDR31}. The earlier data release (\textit{Gaia} EDR3) provided celestial positions (right ascension, $\alpha$ and declination, $\delta$) and the apparent brightness in the $G$-band for $1.8$ billion sources and for $1.5$ billion of those sources, parallaxes, proper motions, and the blue-red photometry ($G_{\rm BP}$-$G_{\rm RP}$) colors were also published. \textit{Gaia} DR3 therefore contains some 585 million sources with five-parameter astrometry (two positions, the parallax, and two proper motion components), and about 882 million sources with six-parameter (6-p) astrometry, including an additional pseudocolour parameter \citep{GaiaEDR31}. The third data release is also accompanied by the astrophysical parameters data which were produced by the Astrophysical parameters inference system (Apsis) based on mean $BP$/$RP$, spectra from Radial Velocity Spectrometer (RVS), astrometry and photometry \citep{APSIS}.

The Astrophysical parameters table contains The Final Luminosity Age Mass Estimator, \textit{FLAME}, which aims to produce the stellar mass and evolutionary parameters for each \textit{Gaia} source that has been analyzed by General Stellar Parametrizer from Photometry (GSP-Phot) and/or General Stellar Parametrizer from Spectroscopy (GSP-Spec). The \textit{FLAME}~parameters include the mass $M$, age $\tau$, evolutionary stage $\epsilon$ along with other parameters. \textit{FLAME}~uses as input data $T_{\rm eff}$, $\log_{10}[g]$, and metallicity $[\rm{M}/\rm{H}]$ from GSP-Phot best data and, when available, these same parameters from GSP-Spec Matisse-Gauguin, along with a distance estimate, $G$-band photometry, and extinction from GSP-Phot. To infer $M$, $\tau$, and $\epsilon$, the BaSTI5 \citep{BASTI5} solar-metallicity stellar evolution models are employed, which consider a mass range from $0.5$-$10$ \(\textup{M}_\odot\) and evolution stage from Zero-Age Main-Sequence (ZAMS) until the tip of the red giant branch \citep{APSIS}. 

\noindent The data that we use in this paper, contains astrometric solutions and the cross matched astrophysical parameters of main-sequence and giant stars in the solar neighborhood. We review and discuss the implemented filtering conditions and properties of this data set in the following subsections.

\subsection{Data Filtering and Conditions} \label{subsec:Filter}
Although the astrometric and photometric measurements in \textit{Gaia} DR3 are highly accurate, there are still some necessary data filtering and cleaning conditions that need to be assessed in the data for having a reliable outcome.

The Renormalised Unit Weight Error, \texttt{ruwe} in the \textit{Gaia} Archive, is defined as an indicator for the source multiplicity (see: \citet{RUWE}) and is expected to be around $1.0$ for sources where the single-star model provides a good fit to the astrometric observations. A value significantly greater than $1.0$ could indicate that the source is non-single or else problematic for the astrometric solution. Hence, we only consider sources where $\texttt{ruwe}\leq1.4$.

Now, we consider more conditions that are similar to the filtering method used in \citet{HRD2018} for the main-sequence stars. It is reported by \citet{Lindegren2021} that the median uncertainty for parallax of bright sources ($G\leq14$ mag) is $0.03$ mas. The systematic errors are lower than $0.1$ mas, and the parallax zeropoint error is about $0.03$ mas. For the photometric solutions, \citet{Evans2018} reports that the precision at $G=12$ is around $1$ mmag in the three passbands, with systematic error at the level of $10$ mmag. For \textit{Gaia} observations, a five-parameter solution is accepted only if at least six visibility periods are used (e.g., the number of groups of observations separated from other groups by a gap of at least four days, this parameter can be found as \texttt{visibility\textunderscore periods\textunderscore used} in the \textit{Gaia} archive). The observations need to be well spread out in time to provide reliable five-parameter solutions. So, we have applied the following filter on this
parameter: $\texttt{visibility\textunderscore periods\textunderscore used}\geq8$ which leads to the removal of strong outliers.

We also adopt a $10 \%$ relative parallax precision criterion, which corresponds to an uncertainty on magnitude in the $G$-band, $\sigma_{G}$, smaller than $0.22$ mag: $\texttt{parallax\textunderscore over\textunderscore error} > 10$. And for removing the variable stars in our data, we should consider filters on the relative flux error on the
G, $G_{\rm BP}$, and $G_{\rm RP}$ photometry: $\texttt{ phot\textunderscore g\textunderscore mean\textunderscore flux\textunderscore over\textunderscore error} > 50$ ($\sigma _G < 0.022$ mag), $\texttt{ phot\textunderscore rp\textunderscore mean\textunderscore flux\textunderscore over\textunderscore error} > 20$, and
$\texttt{phot\textunderscore bp\textunderscore mean\textunderscore flux\textunderscore over\textunderscore error} > 20$ ($\sigma_{GXP} < 0.054$ mag).

After cleaning the data, we need to implement some criterion to limit the data to the sources in the solar neighborhood. For this purpose the following conditions are used: $\texttt{parallax}\geq6.7$ mas (distance $\leq150$ parsecs) and $\texttt{phot\textunderscore g\textunderscore mean\textunderscore mag}\leq21$ mag. 

All of these stars have similar global velocities to the Sun. We note that the stellar global velocity depends only on the distance from the Galactic center as given by \citep[see, e.g.,][]{,2009Rahal,2017Moniez,2019ApJEilers}:
\begin{eqnarray}
v_{\rm rot}(R)= v_{\rm rot, \odot} \Big(1.00762 \big(\frac{R}{R_{\odot}}\big)^{0.039} +0.00712\Big), 
\end{eqnarray}
\noindent where, $R$ is the radial distance from the Galactic center, $v_{\rm rot, \odot}= 220~\rm{km~s^{-1}}$ is the solar rotational velocity, and $R_{\odot}$ is the Sun's distance from the Galactic center \citep{2012Schonrich}.

We also note that since FLAME does not calculate mass and age for all of the stars, we only take the sources where mass and age measurements exist.
\begin{figure}[]
	\subfigure[]{\includegraphics[angle=0,width=0.49\textwidth,clip=]{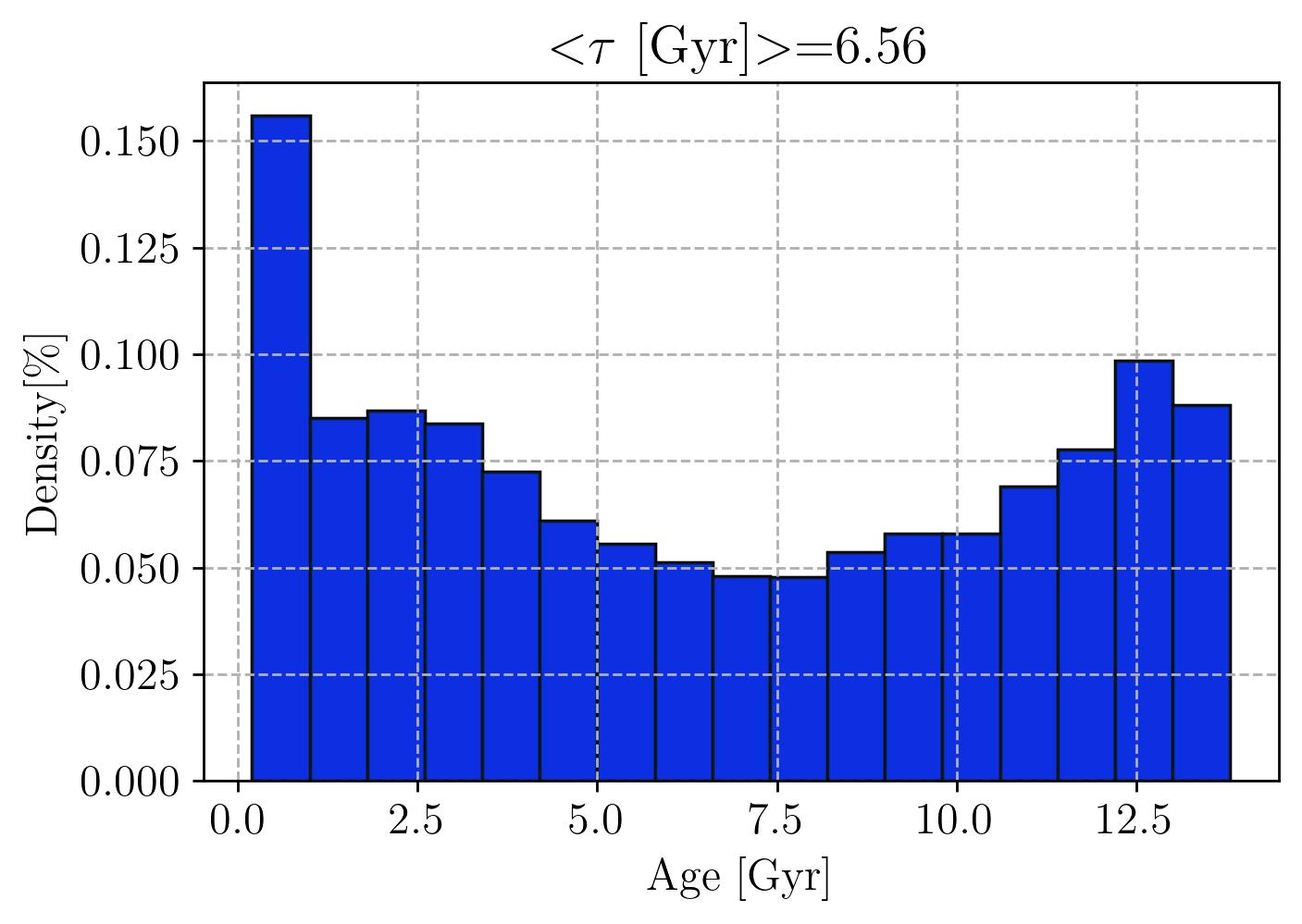}\label{age}}
	\subfigure[]{\includegraphics[angle=0,width=0.49\textwidth,clip=]{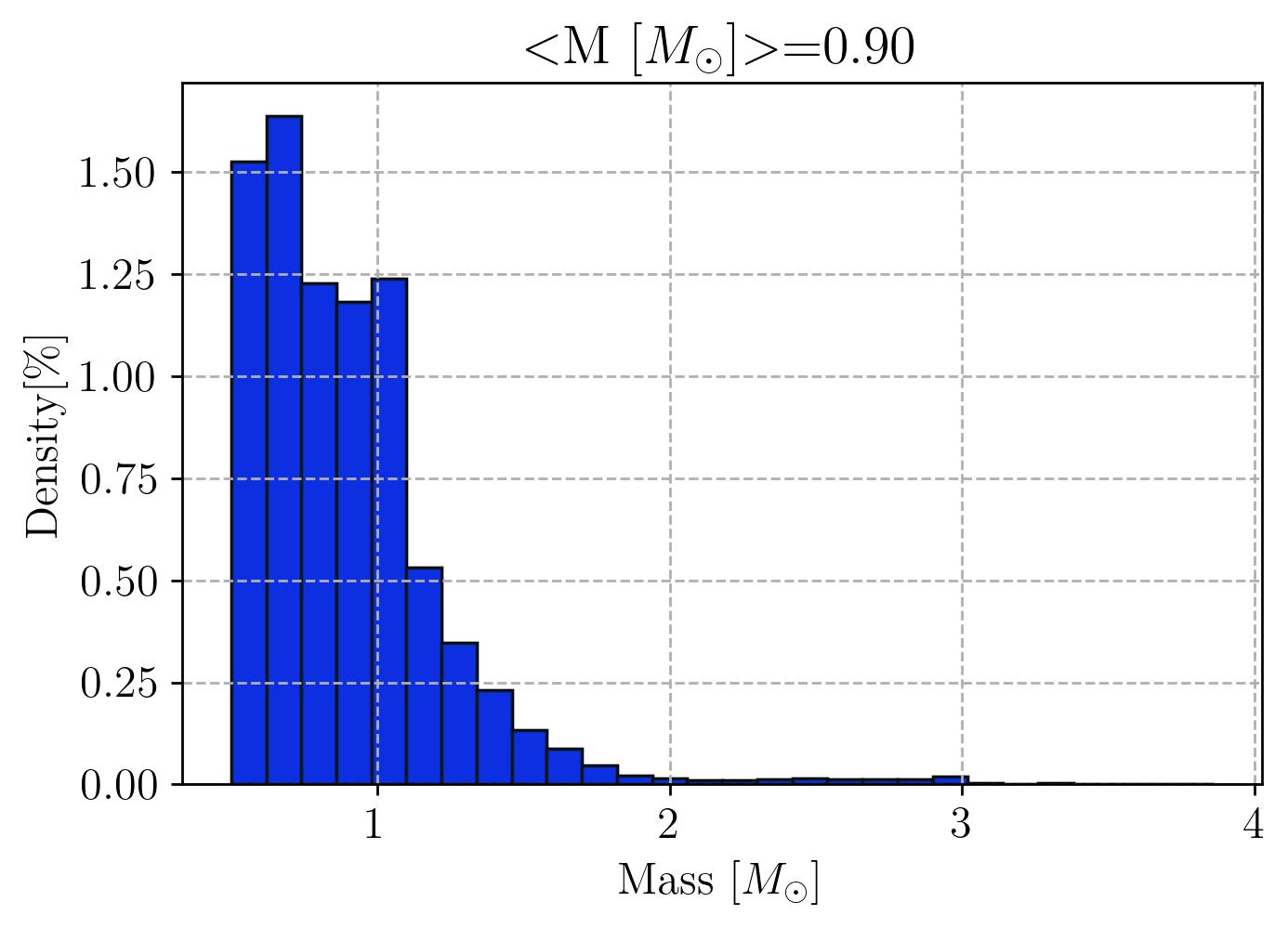}\label{mass}}	
	\subfigure[]{\includegraphics[angle=0,width=0.49\textwidth,clip=]{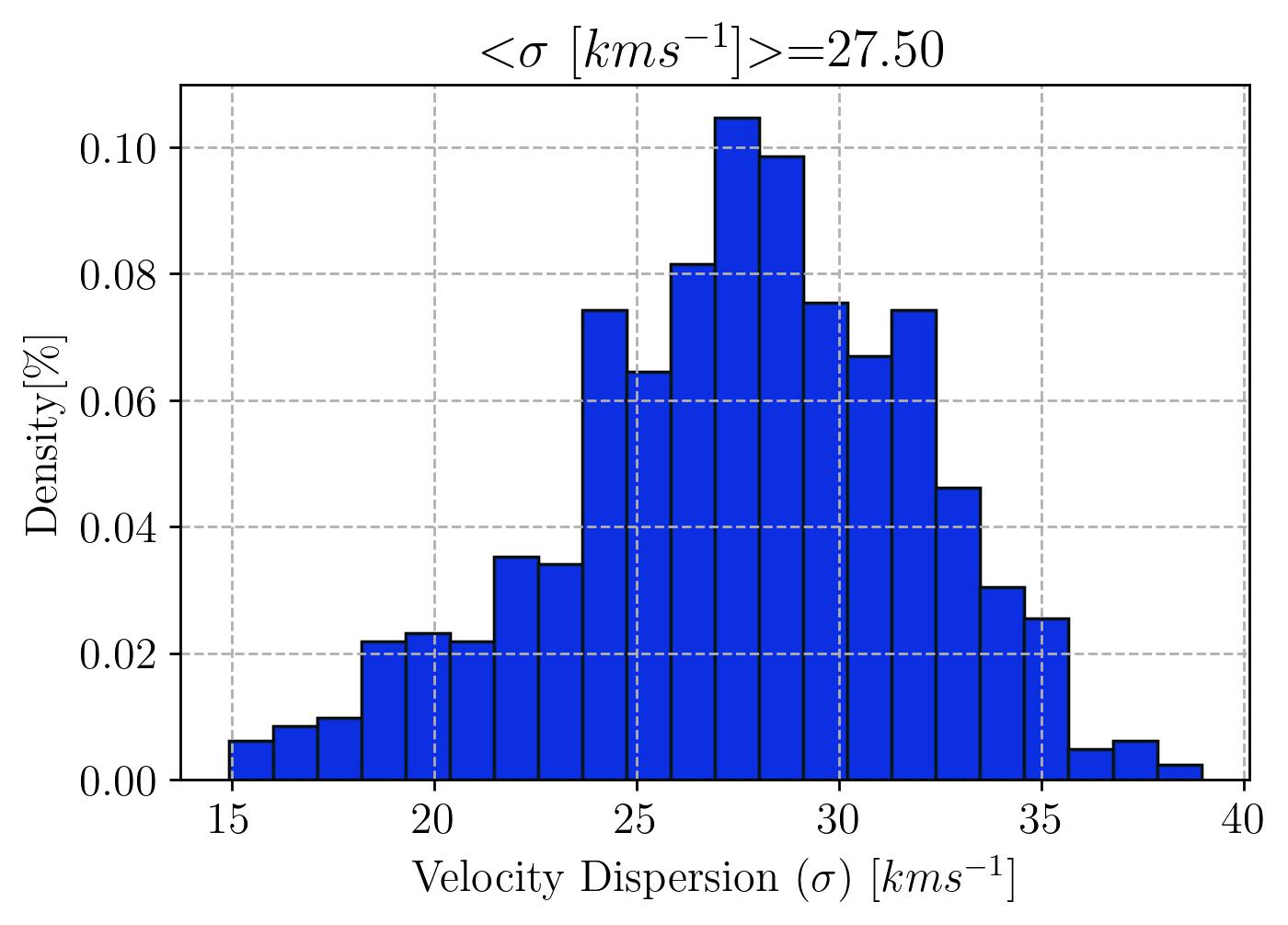}\label{sigma}}		
	\caption{The normalized histograms of the stellar age, mass, and velocity dispersion of the stars in our sample chosen from GDR3, from top to bottom panel, respectively.}
	\label{fig:hist}
\end{figure}

\subsection{Feature Engineering} \label{subsec:feature}
In this subsection, we explain our methods for preparing the data set for statistical learning and modeling. The data set is divided into two subsets; the predictor (input) variables and response (output) variable:    

{\bf Predictor variables:} The predictors in our analysis are the stellar mass and age. The reported mass (normalized to the solar mass) and age (Gyr) values in Astrophysical parameters table are calculated in the $16^{\rm th}$, $50^{\rm th}$ (the median value) and $84^{\rm th}$ percentiles of the 1D projected distribution from sampling in mass and age \citep{APSIS}. Since, by definition, there is a $68\%$ chance that the correct values of these parameters fall between the $16^{\rm th}$ and $84^{\rm th}$ percentiles, we use the $50^{\rm th}$ percentile (the median). The histograms of these variables are illustrated in Figures \ref{age}, and \ref{mass}. Accordingly, our sample contains stars in wide ranges of age ($[0.2,~13.5]$ Gyrs, containing both main-sequence and giant stars) and mass ($[0.5,~3.8] M_{\odot}$).

{\bf Response variable:} The response variable of our Machine learning analysis is the stellar velocity dispersion, $\sigma$ which are calculated according to the velocity components reported in the \gaia~DR3, i.e., $\big(v_{\rm ra},~v_{\rm dec},~v_{\rm r}\big)$, which are projected velocity components on the sky plane in the right ascension and declination, and its radial component in the line of sight direction. The reference epoch of all measurements in \textit{Gaia} DR3 is $J2016.0$. 

We note that the proper motions and radial velocities are measured in the International Celestial Reference System (ICRS), i.e., heliocentric frame. Before we can do any preliminary analysis of the data, we need to transform the velocity components from the heliocentric frame to the Local Standard of Rest (LSR) frame, and remove the peculiar solar velocities from the measured velocities (see appendix \ref{ap:VTransform}). For this purpose, proper motions in each given right ascension and declination directions, as well as radial velocities are first transformed into Galctocentric frame, in which the velocity components are represented as $(v_{\rm U},~v_{\rm V},~v_{\rm W})$.

Finally, we consider the absolute velocity of each star (in LSR) as: $v=\sqrt{v_{\rm U}+v_{\rm V}+v_{\rm W}}$. We note that the rotation of the coordinate systems do not change this size, so after subtracting the solar peculiar velocities in the Galactic frame, we do not need to convert them back to the heliocentric frame.   

The Statistical Dispersion is a measure of describing the extent of distribution of a group of data around a central value or point. So, we need to equally divide our data set into subgroups in order to calculate Velocity Dispersions. After sorting the data with respect to mass, we then equally divide our data set into 753 subgroups, with each containing 150 sources. For these subgroups, mass and age averages are calculated. If the subgroup velocities are further divided into quartiles, or four rank-ordered even parts via linear interpolation, where the quartiles are denoted by $Q_1$ (the $25\rm{th}$ percentile), $Q_2$ (the median), $Q_3$ (the $75\rm{th}$ percentile) and $Q_4$ (the upper quartile), we can choose the interquartile range (IQR) between the velocities as the Velocity Dispersion ($\sigma$) for each subgroup as the following \citep[e.g., ][]{Manikandan2011,2010Brown}:
\begin{equation}
\mu = Q_3 - Q_1.
\end{equation}

\noindent The IQR provides a highly reliable descriptive statistic as a measure of the Dispersion in the stellar velocities since it is not affected by extreme values in each subgroup. Figure \ref{sigma} illustrates normalized histogram plot of the stellar velocity Dispersions of our samples, as the response variable. The average value is mentioned on top of the plot.

Before we can model our data, we need to split the data into two subsets; a training set and a test set. The model is going to be trained based on the training set, and the accuracy of the model would be assessed over the test set. Since the test set is not used in the process of modeling the data, we can examine how well our model can make predictions over a group of data that it has never acquired, by comparing the model responses to the actual response values in the test set. Since our data set is not large enough, choosing how to divide the data into these subgroups could highly effect the accuracy measurements and the results may vary accordingly. To solve this problem, we implement a data resampling method called \textit{K-Fold Cross-Validation}.

\subsection{K-Fold Cross-Validation}
K-Fold Cross-Validation, involves randomly sorting the set of observations into k groups or \textit{folds}, where each fold has the same amount of data as the original set, but in a random order, and in each fold, randomly split the data into a training set and a test set. This results in $k$ estimates of the Root Mean Squared Error (RMSE) as $\rm{RMSE}_1, \rm{RMSE}_2, ..., \rm{RMSE}_k$. $\rm{RMSE}_i$ is RMSE of the $i$th fold which is given by: 
\begin{equation}
\rm{RMSE}_i=\sqrt{\frac{1}{N}\sum_{j=1}^N \Big(\left<y_i\right>-y_j \Big)^2},
\label{eq:RMSE}
\end{equation}  

\noindent where, $\left<y_i\right>$ is the average response value in the $i\rm{th}$ fold, $y_j$ is the predicted values from the model, and $N$ is the total number of test data in each fold. The final model accuracy can simply be defined as:
\begin{equation} \label{eq:CV}
E=\frac{1}{k} \sum^{k}_{i=1}\rm{RMSE}_i .
\end{equation}  

For our data set, which contains 753 rows of data, we re-sample the data into 5 folds ($k=5$), with each fold containing about $N=602$ rows ($80\%$) of data as the training set and $M=151$ rows ($20\%$) as the test set.

Now that our training and test sets are fully prepared, we can model our data. The model and our methods are explained in the next section.

\begin{figure*}[]
	\subfigure[]{\includegraphics[angle=0,width=0.49\textwidth,clip=]{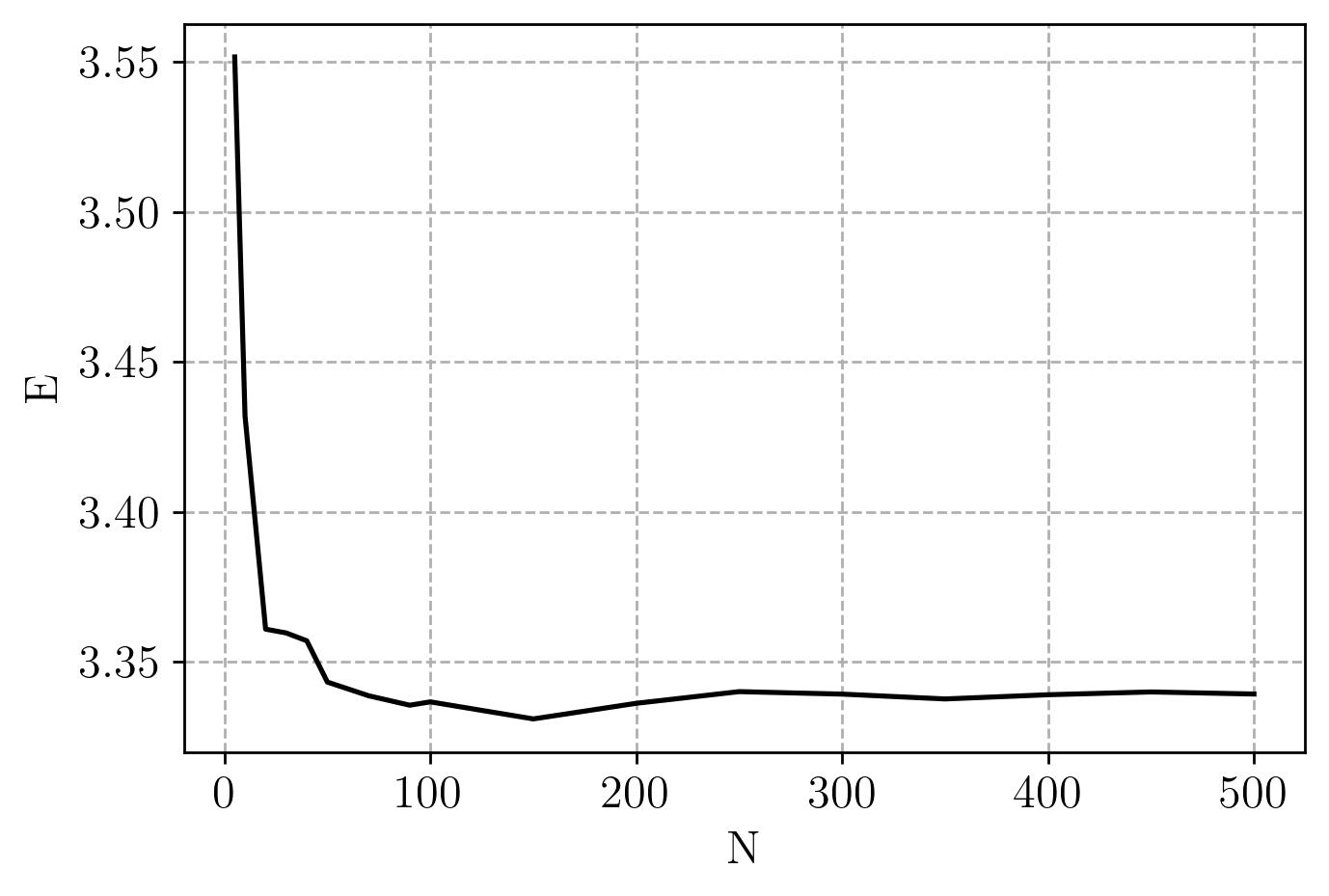}\label{fig1}}
	\subfigure[]{\includegraphics[angle=0,width=0.49\textwidth,clip=]{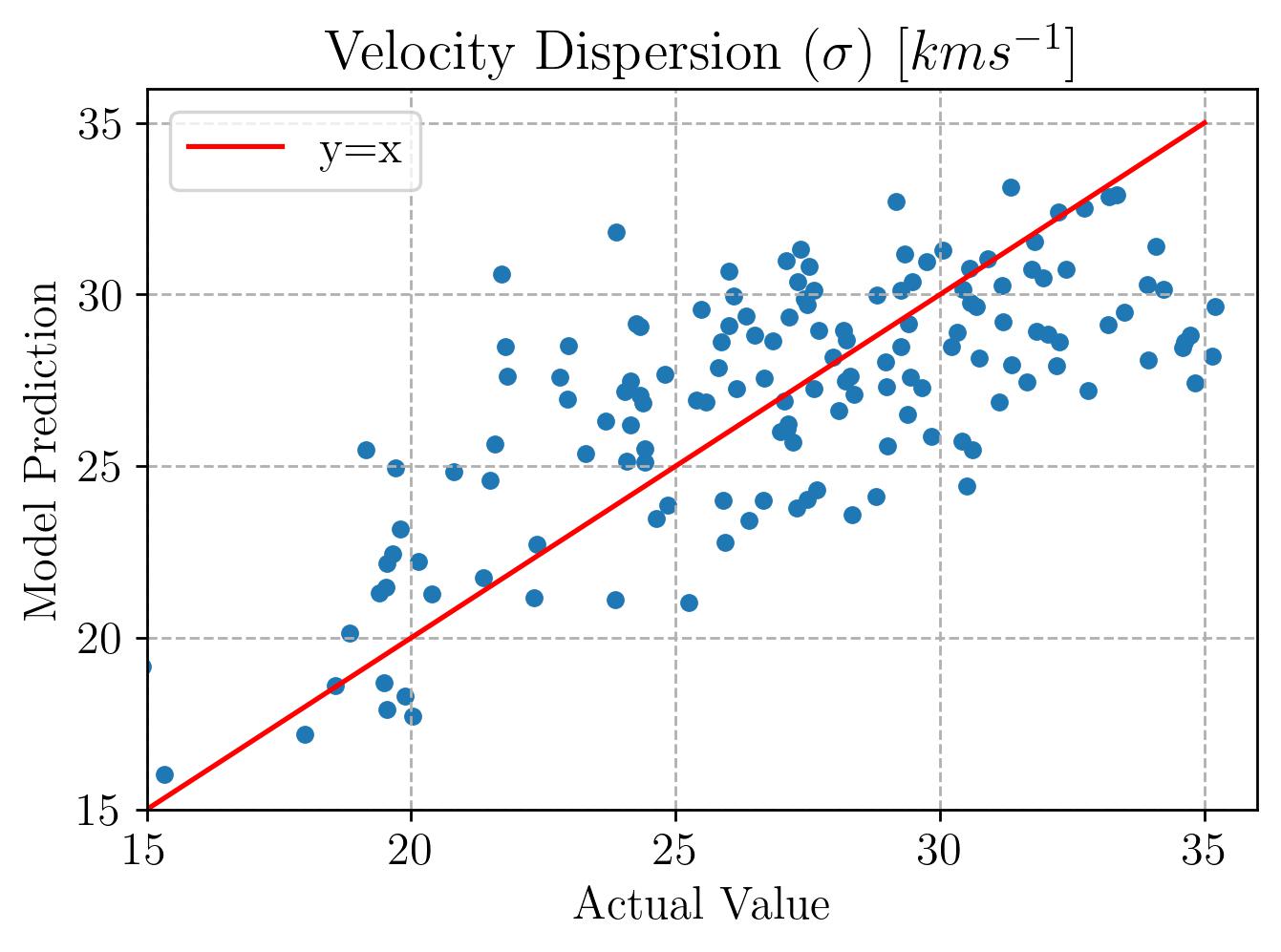}\label{fig2}}		
	\caption{Left panel: final error (eq \ref{eq:CV}) for different numbers of trees, and right panel: Velocity-Dispersion prediction versus actual value for one of the test sets. \label{fig:results}}
\end{figure*}

\section{Methodology} \label{sec:Method}
In this section, we review and explain the algorithms that were implemented for constructing our statistical model over the data set. Our model is mostly based on an ensemble of tree-based methods, which is referred to as \textit{Random Forests}, where the k-folds algorithm is used for the model cross-validation. We have chosen this machine learning algorithm for its high interpretability, accurate predication and most importantly for its ability to assess features correlation and relative importance \citep{vilone2020}.   

Tree-based methods \citep{Quinlan86inductionof}, or decision trees, are one of the most useful supervised statistical learning methods \citep{hastie2009elements}. Tree-based algorithms are helpful predictive models with high accuracy, stability and are easy to interpret. Unlike their linear counterparts, they can also map non-linear relationships quite well. This method makes predictions by dividing the feature space (the predictor variables) in a way that the data is split recursively into n distinct and non-overlapping regions, $R_1, R_2,...,R_n$ and for every observation that falls in the region $R_i$, the mean of the response values in that region is considered as the prediction \citep{james2013introduction}. 

\noindent The predictor space is divided into regions (in the form of N-dimensional intervals) with the goal of minimizing the Residual Sum of Squares (RSS), which is defined as:
\begin{equation} \label{eq:RSS}
\rm{RSS}=\sum^{n}_{i=1}\sum_{j \in R_i}\Big(\left<y_{\rm{Ri}}\right>-y_j\Big)^2 ,
\end{equation}
where $\left<y_{\rm{Ri}}\right>$ is the mean of response values for the training set in the $i_{th}$ region and $y_j$ is the actual value.

For minimizing the equation \ref{eq:RSS}, \textit{Recursive Binary Splitting} (RBS) is used. In RBS we begin splitting the data at the top of the tree, where all observations are in one single region, and then successively split the predictor space into binary regions at a time. Each split of the data is represented as a \textit{branch}, where the \textit{node} is the decision that was made for the splitting, and hence the model is called a \textit{Decision Tree}. 

\noindent The RBS process first selects a predictor variable, $X_i$ and the border value s, in a way that dividing the predictor space into the regions $R_1(i,~s)=\{X|X_i < s \}$ and $R_2(i,~s)=\{X|X_i > s \}$ results in the greatest possible reduction in RSS (equation \ref{eq:RSS}). 

\noindent Now, we can look for the value of i and s that minimize $\mathcal{F}$ as defined in the following:
\begin{equation} \label{eq:RSS2}
\mathcal{F}=\sum_{j:x_j\in R_1(i,s)}\Big(\left<y_{R1}\right>-y_j\Big)^2 +  \sum_{j:x_j\in R_2(i,s)}\Big(\left<y_{R2}\right>-y_j\Big)^2 ,
\end{equation}
where $\left<y_{R1}\right>$ and $\left<y_{R2}\right>$ are the average response values in the training observations in $R_1(i,s)$ and $R_2(i,s)$, respectively.

\noindent When two regions are made based on this algorithm, each resulting region is further split into two other regions and the process is repeated until a stopping thresh-hold is reached i.e. when there are a certain number of data in each region \citep{james2013introduction}.

The RBS procedure forces the Decision Trees algorithm to be highly dependent on the input features and there are some instances where these models tend to overfit the data, i.e. when a very small stopping criterion is considered, the decision tree tries to becomes so accurate that it even models the outliers and errors in the training set, and when it wants to make a prediction for a data which is not present in the training set, it fails to produce an accurate response. 

\noindent One way that we can resolve these problems is by using \textit{Random Forests} (RF) \citep{breiman2001random}. RF enhances the predictive performance of decision trees, substantially. 

\noindent RF is an ensemble approach that uses and combines many repeated models (which are referred to as \textit{weak learners or models}) in order to produce a single and powerful model by averaging over the responses of various single models.

One important factor is trying to train each weak learner differently, and each time with a different training subset, because failing to do so, would generate many correlated single trees that produce the same response variable.

\noindent To avoid this problem we train different trees in RF based on \textit{Bootstrapping} the data sets. Bootstrapping \citep{efron1979computers} is a statistical data resampling procedure that uses a single data set to create many simulated samples. This process creates random samples from the original data set in a way that an equal probability exists for randomly drawing each original data point for inclusion in the resampled data sets. It can also select a data point more than once for each resampled data set. In this way, the bootstrapping is able to produce resampled data sets that are the same size as the original data set \citep{kulesa2015sampling}. In RF, after bootstrapping the training set, N number of separate decision trees are trained which produces $\left<f^1(X)\right>,~\left<f^2(X)\right>,~...,~\left<f^N(X)\right>$, where $\left<f^i(X)\right>$ is the result of the $i^{th}$ tree. The final response can be easily calculated as:
\begin{equation} \label{eq:RF1} 
\left<f_T(X)\right>=\frac{1}{N}\sum_{i=1}^{N}\left<f^i(X)\right>.
\end{equation}   
The accuracy of our model depends on the choice of N, only to some extent. If we take a relatively high number for N, more trees are trained, and we would have a more accurate model in return. But, after a while, increasing N would not affect the resulting final RMSE (equation \ref{eq:CV}), because we have reached the best possible model.

For implementing the models and algorithms discussed in this section, we use the \texttt{PYTHON}-based open-source machine learning package, \texttt{Scikit-Learn} \citep{scikit-learn}. The results of this model are reported in the next section.

\begin{deluxetable}{cccccc}
\tablenum{1}
\tablecaption{Model Precision Measurements. \label{tab:Errors}}
\tablewidth{0pt}
\tablehead{\colhead{N}&\colhead{\rm{MSE}}&\colhead{\rm{RMSE}}&\colhead{\rm{N}-\rm{RMSE}}&\colhead{MAE}&\colhead{N-MAE}\\
\colhead{} & \colhead{$\rm{(km^{2}s^{-2})}$} & \colhead{$\rm{(kms^{-1})}$} & \colhead{} & \colhead{$\rm{(kms^{-1})}$} & \colhead{}}
\startdata
100 & 10.708 & 3.272 & 0.119 & 2.682 & 0.097 \\
150 & 10.694 & 3.270 & 0.119 & 2.679 & 0.097 \\
200 & 10.706 & 3.272 & 0.119 & 2.684 & 0.097
\enddata
\end{deluxetable}
\begin{figure}[]
\subfigure[]{\includegraphics[angle=0,width=0.49\textwidth,clip=]{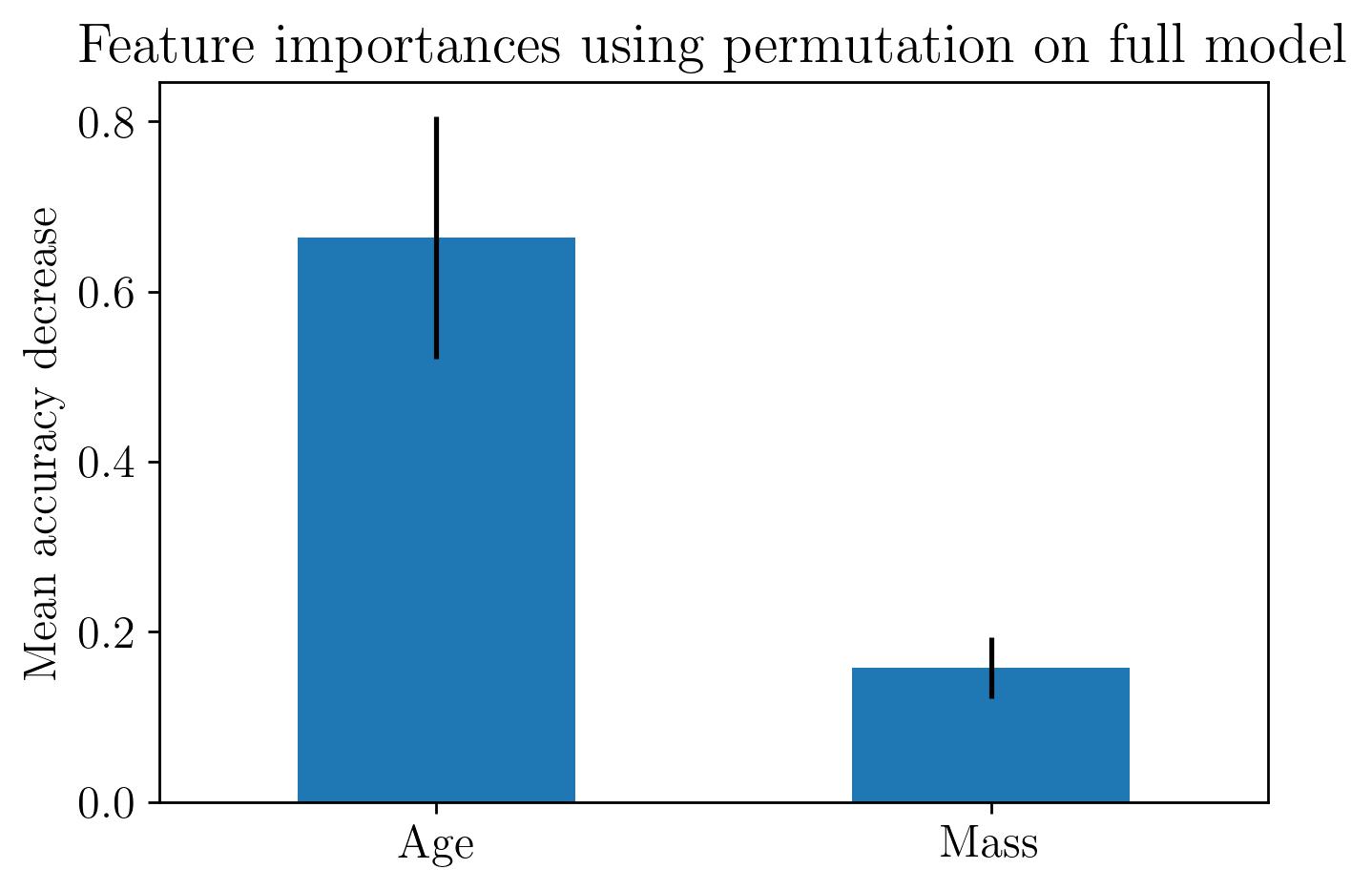}\label{fig3a}}
\subfigure[]{\includegraphics[angle=0,width=0.49\textwidth,clip=]{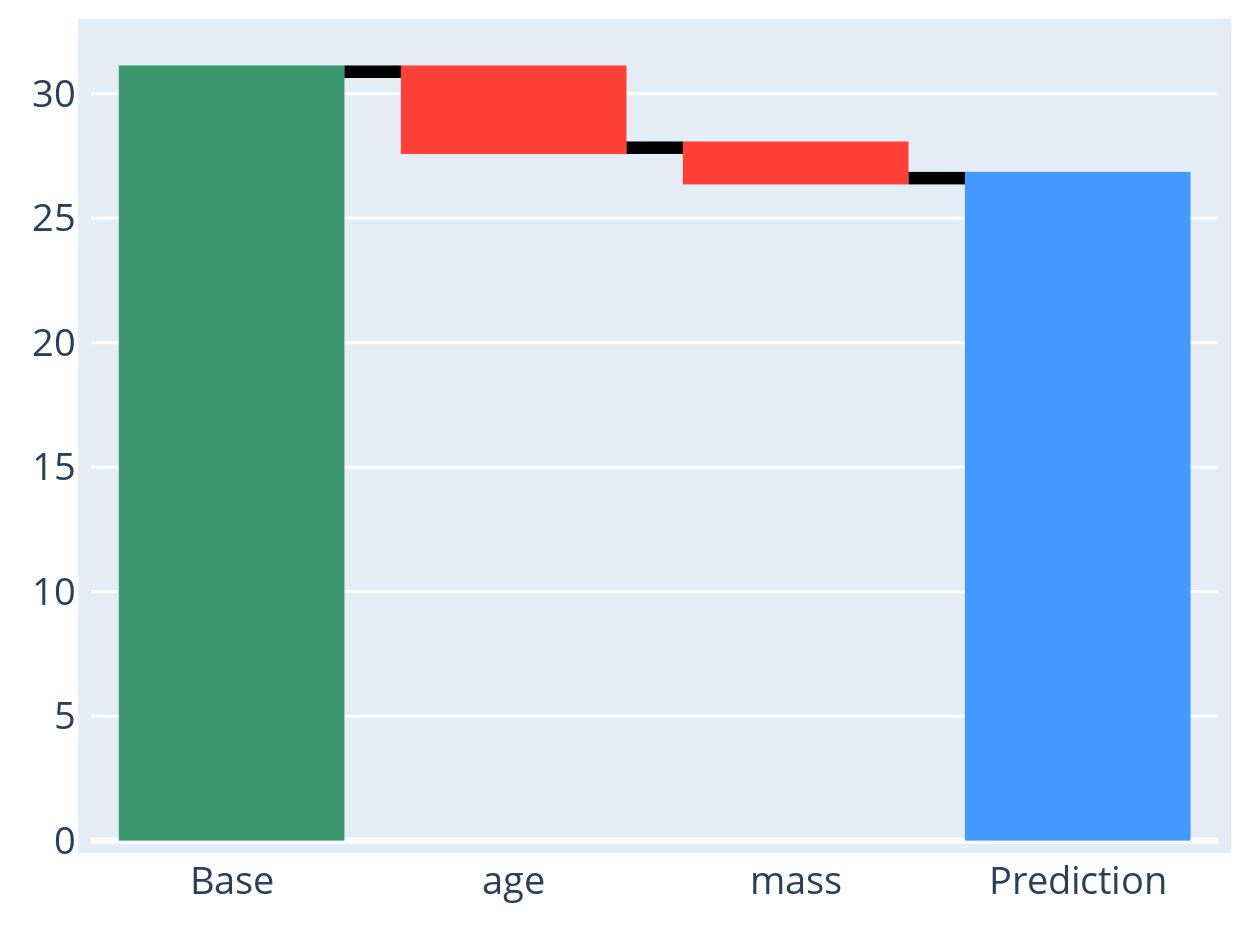}\label{fig3b}}			
\caption{Top panel: Final Model Feature Importance. Bottom panel: Mean Decision Making Procedure for a random sample. \label{fig:importance}}
\end{figure}

\section{Model Results and Analysis} \label{sec:Results}

First we need to find the appropriate number of weak learners, or \textit{trees}, (N) in our model (a single weak learner which only has a depth with the order of 4 is shown in Appendix \ref{ap:treedig}). For this purpose, we plot the final RMSE (equation \ref{eq:RMSE}) and the Number of trees in figure \ref{fig1}. As we can see, RMSE reaches its minimum, $3.60 \rm{km s^{-1}}$ in $N \in [100,~200]$ and then slightly rises afterwards.

In Table \ref{tab:Errors} we report averages of Mean Squared Errors (MSE), Root Mean Squared Error (RMSE), Normalized RMSE (N-RMSE), Mean Absolute Error (MAE) and Normalized MAE (MAE) for 3 different number of trees (N). The MAE is defined as:

\begin{equation} 
\rm{MAE}_i=\frac{1}{n}\sum_{j=1}^n \Big| \left<y_i\right>-y_j \Big|.
\label{eq:MAE}
\end{equation} 

Since RMSE is minimized at $N=150$, we choose $N=150$ as the number of trees in our model. For this number of trees, and one of the randomly chosen cross-validation folds, we plot the model prediction vs the actual value in Figure \ref{fig2}. The closeness of dots to the identity line ($y=x$) shows the model accuracy.

As we mentioned in section \ref{sec:Data}, our predictor space consists of Mass (M) and Age ($\tau$). In RF, there is no need to try different models in order to assess the importance of each predictor variable. Instead, this can be done via a \textit{Permutation Based} approach \citep{Altmann2010}.

\noindent The Permutation Based method works by randomly shuffling each feature and computing the change in the model's performance, based on its accuracy. The feature which impacts the performance more, is then considered to be of more importance.

\noindent Figure \ref{fig3a} displays the correlation parameter or importance of Age and Mass in the final model; we can see that, on average, the Age feature is between $3$ to $8$ times more important than Mass in our model. But it also should be noted that the exact value depends on the stellar type and mass. As was expected, Age is a more important feature for calculating the Velocity-Dispersion. 

Figure \ref{fig3b} shows the contributions of each feature to the decision making procedure. A sample containing 60 data was randomly chosen, where the average of actual target values (velocity-dispersion) was about $26.86$ $\rm{kms^{-1}}$. The model predicted an average velocity-dispersion of $26.61$ $\rm{kms^{-1}}$. The RF started with a base value of $30.883$ $\rm{kms^{-1}}$ (based on the predictor space of the sample). On average, Age and Mass parameters lead to a deduction of $3.052$ $\rm{kms^{-1}}$ and $1.22$ $\rm{kms^{-1}}$ respectively. Hence the predicted value of $26.61$ $\rm{kms^{-1}}$ was reached.

\section{Conclusions}\label{Conclusions}
Different interactions of stars with giant structures in our Galaxy leads to the evolution of stellar random velocities with time. Accordingly, velocity dispersions increase with time, which has been known as Age-Velocity Dispersion Relation (AVR). There are some other weak interactions of stars with different Galactic components that cause a second-order dependence of AVR with stellar mass. For instance, interaction of stars with MACHOs or other stars. In this work, we used the Gaia data archive of stars in the Solar neighborhood to find a three-fold relation between mass, age, and velocity dispersion of stars, i.e., Mass-Age-Velocity dispersion relation (MAVR). For a big ensemble of these stars with reported mass, age, and velocity components contains 113035 ones, we tried to find MAVR using Random Forest (RF) numerical approach in Supervised Machine Learning.  

We first divided the initial sample of stars into 753 subsamples with different average mass and age values, and then determined velocity dispersion of stars in each subsample. Then, we applied K-Fold cross validation (with k=5) of Random Forest Regressors (each with 100-200 trees) to subsamples. Our model can determine velocity dispersion as a function of stellar mass and age with Mean Absolute Error of $2.68~\rm{km/s}$. We note that the average stellar velocity of stars in our initial sample was $27.50~\rm{km/s}$ (see Figure \ref{fig:hist}), which results the relative error of about $9\%$. For our data sample and model, the importance of stellar age while calculating velocity dispersion is $3$ to $8$ times more than that due to stellar mass.

The raw queried data can be found at:
\href{https://github.com/AstroFatheddin/Gaia-MAV-Data}{https://github.com/AstroFatheddin/Gaia-MAV-Data}

All of the Machine learning algorithms discussed in this paper are available at:
\href{https://github.com/AstroFatheddin/GaiaMAV}{https://github.com/AstroFatheddin/GaiaMAV}

\section*{Acknowledgements}
H. F. thanks Mozafar Allahyari for his help and support. This work has made use of data from the European Space Agency (ESA) mission {\it Gaia} (\url{https://www.cosmos.esa.int/gaia}), processed by the {\it Gaia} Data Processing and Analysis Consortium (DPAC, \url{https://www.cosmos.esa.int/web/gaia/dpac/consortium}). Funding for the DPAC has been provided by national institutions, in particular the institutions participating in the {\it Gaia} Multilateral Agreement. We also thank the reviewer for his/her useful comments and suggestions.

\newpage
\appendix
\section{Velocity Transformations} \label{ap:VTransform}
Here we explain the velocity transformations from ICRS to LSR (Galactic Coordinates).
\\The ICRS coordinates of the north galactic pole are $(\alpha_ G,\delta_G)=(192.85948^{\circ},+27.12825^{\circ})$ \citep{HIPPARCOS1997} and the galactic longitude of the first intersection of the galactic plane with the equator is $l_{\Omega}=32.93192^{\circ}$. First we write ICRS velocity components in ICRS unit vectors:
\[
\begin{bmatrix}
    VX_{\rm ICRS}  \\
    VY_{\rm ICRS}  \\
    VZ_{\rm ICRS} 
\end{bmatrix} 
=\begin{bmatrix}
    cos(\alpha)  \\
    -sin({\alpha})  \\
    0 
\end{bmatrix} \times \mu_{\alpha}
+ \begin{bmatrix}
    -sin(\alpha)sin(\delta)  \\
    -cos(\alpha)sin(\delta)  \\
    cos(\delta)
\end{bmatrix} \times \mu_{\delta}
+ \begin{bmatrix}
    cos(\delta)  \\
    cos(\delta)  \\
    sin(\delta)
\end{bmatrix} \times v_{r}
\]
\\ Where $\mu_{\alpha}$ and $\mu_{\delta}$ are proper motion components in right ascension and declination directions and $v_r$ is the radial velocity.
\\ In terms of the column matrices, the transformation from ICRS to the galactic system can be easily obtained through the matrix multiplication:
\[
\begin{bmatrix}
    VX_{\rm GAL}  \\
    VY_{\rm GAL}  \\
    VZ_{\rm GAL} 
\end{bmatrix} 
= \begin{bmatrix}
    VX_{\rm ICRS}  \\
    VY_{\rm ICRS}  \\
    VZ_{\rm ICRS}
\end{bmatrix} \times A_{G}^{\prime}
\]
\\where $A_{G}^{\prime}$ is a fixed orthogonal matrix, defined as: \citep{HIPPARCOS1997}

\[
A_{G}^{\prime}=R_z(-l_{\Omega})R_x(90^{\circ} - \delta _G)R_z(\alpha _G + 90^{\circ})
=\begin{bmatrix}
    -0.0548755604162154       & -0.8734370902348850 & -0.4838350155487132 \\
    +0.4941094278755837       & -0.4448296299600112 & +0.7469822444972189 \\
    -0.8676661490190047       & -0.1980763734312015 & +0.455983776175066 
\end{bmatrix} 
\].
More information, see: \href{https://gea.esac.esa.int/archive/documentation/GDR2/Data_processing/chap_cu3ast/sec_cu3ast_intro/ssec_cu3ast_intro_tansforms.html}{Gaia Website}
\section{A Single Weak Learner Tree Diagram} \label{ap:treedig}
\begin{figure*}[!ht]
\centering
\includegraphics[width=0.99\textwidth]{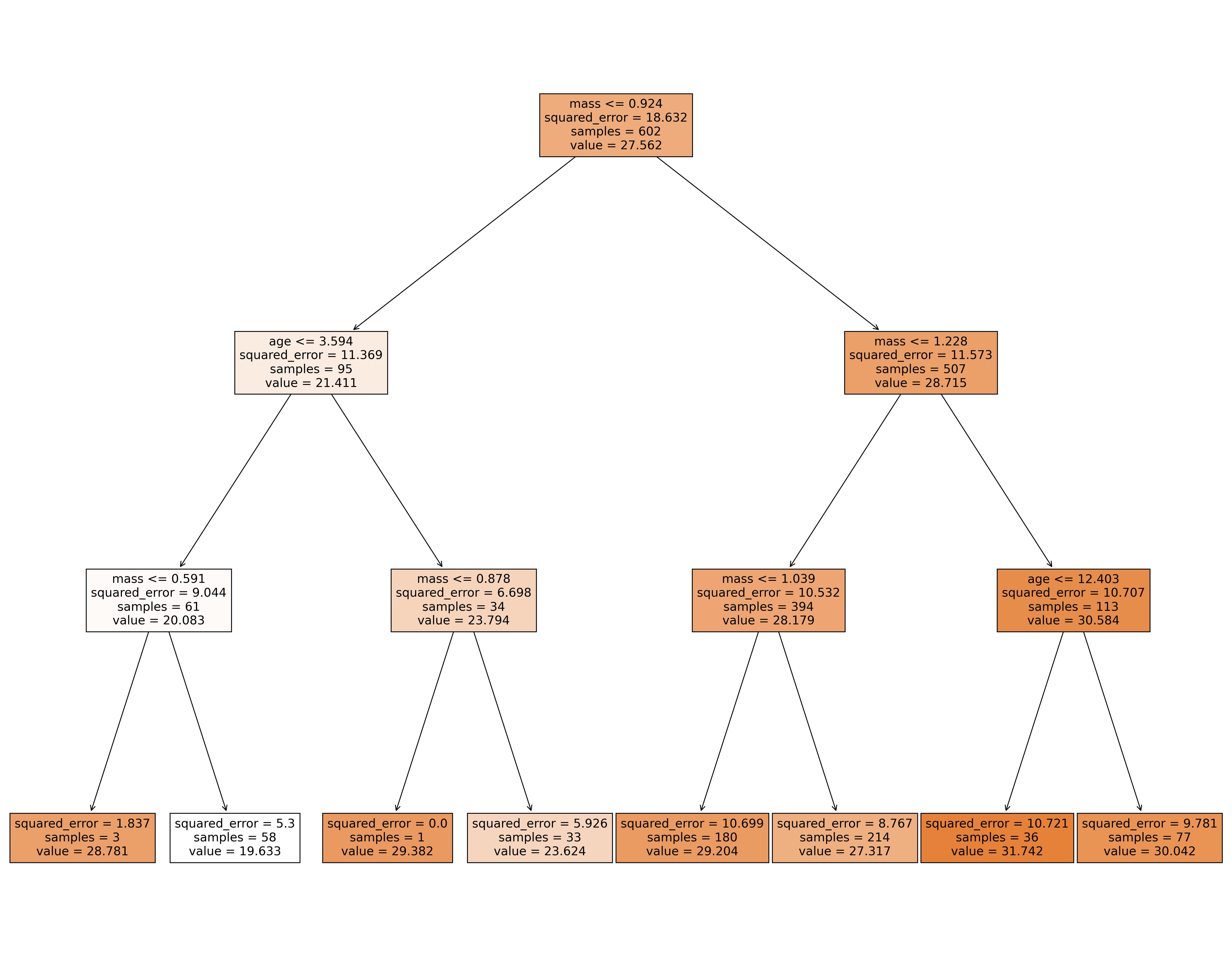}\label{fig:dtr}
	
\caption{Tree Diagram of a small Decision Tree (a weak learner in RF)}
\end{figure*}

Figure \ref{fig:dtr} shows the tree diagram of a Decision Tree over our data set. Each branch in this tree represents a split in the data. Each node (box) shows a splitting condition based on one of the predictors (mass or age). Here we have chosen a tree that only has a growing depth of 3; this means that this tree was only allowed to split the data until $2^3$ regions were produced. Whereas, the weak learners in our final RF model, do not have this condition and can grow until a threshold number of data in each region is reached (these trees can not be illustrated here, due to their relatively large amounts of branches and regions). 

In this diagram, if a condition in a node holds true, the region at the right branch is considered, and vice versa. The boxes at the end of the tree (which are called leaves) contain the squared errors, number of samples and the predicted values.

\bibliography{sample631}{}
\bibliographystyle{aasjournal}
\end{document}